\documentclass{article}
\usepackage[T1]{fontenc} 
\usepackage[utf8]{inputenc} 
\usepackage{ismir,amsmath,cite,url,amssymb,pifont}
\newcommand{\cmark}{\ding{51}}%
\newcommand{\xmark}{\ding{55}}%
\usepackage{graphicx,verbatim}
\usepackage{color}
\usepackage{multirow}
\usepackage[ruled,noend]{algorithm2e}

\usepackage[normalem]{ulem}

\usepackage{lineno}

\title{Supervised Metric Learning for Music Structure Features}

\multauthor
{Ju-Chiang Wang \hspace{0.5cm} Jordan B. L. Smith \hspace{.5cm} Wei-Tsung Lu \hspace{.5cm} Xuchen Song} 
{
ByteDance \\
{\tt\small \{ju-chiang.wang, jordan.smith, weitsung.lu, xuchen.song\}@bytedance.com}
}

\def\authorname{J.-C. Wang, J. B. L. Smith, W.-T. Lu, and X. Song}

\begin{document}

\maketitle
\begin{abstract}

Music structure analysis (MSA) methods traditionally search for musically meaningful patterns in audio: homogeneity, repetition, novelty, and segment-length regularity. Hand-crafted audio features such as MFCCs or chromagrams are often used to elicit these patterns. However, with more annotations of section labels (e.g., verse, chorus, bridge) becoming available, one can use supervised feature learning to make these patterns even clearer and improve MSA performance. To this end, we take a supervised metric learning approach: we train a deep neural network to output embeddings that are near each other for two spectrogram inputs if both have the same section type (according to an annotation), and otherwise far apart. We propose a batch sampling scheme to ensure the labels in a training pair are interpreted meaningfully. The trained model extracts features that can be used in existing MSA algorithms. In evaluations with three datasets (HarmonixSet, SALAMI, and RWC), we demonstrate that using the proposed features can improve a traditional MSA algorithm significantly in both intra- and cross-dataset scenarios.

\end{abstract}
\section{Introduction}\label{sec:introduction}



In the field of Music Structure Analysis (MSA), most algorithms, including many recent and cutting-edge ones~\cite{mcfee2014analyzing, shibata2020music, wang2016structural}, use conventional features such as MFCCs and Pitch Class Profiles (PCPs).
Devising a suitable feature for MSA is challenging, since so many aspects of music---including pitch, timbre, rhythm, and dynamics---impact the perception of structure~\cite{bruderer2009perception}.
Some methods have aimed to combine input from multiple features~\cite{deberardinis2020unveiling}, but this requires care: MSA researchers have long been aware that structure at different timescales can be reflected best by different features (see, e.g.,~\cite{jehan2005a}).

A common story in MIR in the past decade is that using feature learning can improve performance on a task. Although this wave of work arrived late to MSA, we have already seen the benefits of supervised learning to model, for instance, `what boundaries sound like'~\cite{ullrich2014_ismir}, or `what choruses sound like'~\cite{wang2021supervised}. One drawback of these two methods was that they were not compatible with existing MSA pipelines: new post-processing methods had to be conceived and implemented. Also, each one solved a limited version of MSA: segmentation and chorus detection, respectively. Developing a supervised approach that can explicitly minimize the losses of segmentation and labeling tasks at the same time remains a challenge.

In~\cite{mccallum2019unsupervised}, \textit{unsupervised} training was used to create a deep embedding of audio based on a triplet loss that aimed to reflect within-song similarity and contrast.
The embedding vectors, treated as features, can directly replace traditional features in existing MSA pipelines, making it possible to leverage large, unannotated collections for MSA. 
This approach demonstrates the promise of learning features with a deep neural network (DNN) for MSA.

\begin{figure}
\centering
\includegraphics[width=\columnwidth]{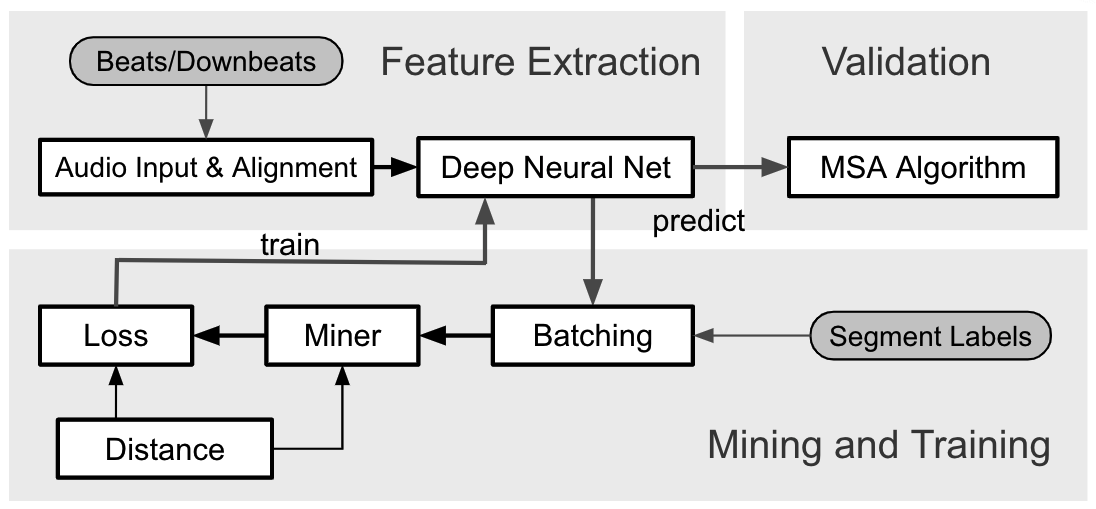}
\caption{The training pipeline.}
\label{fig:sys}
\end{figure}


An unsupervised approach has so far been sensible, given how few annotations exist, and how expensive it is to collect more. However, the appeal of supervised learning has grown with the introduction of Harmonix Set~\cite{nieto2019harmonix}, containing 912 annotated pop songs.
Although Harmonix Set is smaller than SALAMI~\cite{smith2011design} (which has 1359 songs), it is much more consistent in terms of genre, which raises our hopes of learning a useful embedding.
A model trained on SALAMI alone would have to adapt to the sound and structure of pop music, jazz standards, piano concertos, and more; a model trained on Harmonix Set has only to learn the sound and structure of pop songs.
In short, the time is right to pursue a supervised approach.

In this paper, we propose to use supervised metric learning to train a DNN model that, for a given song, will embed audio fragments that lie in different sections far apart, and those from the same section close.
(See Fig.~\ref{fig:sys} for an overview of the training pipeline.)
This approach can help the model to capture the homogeneity and repetition characteristics of song structure with respect to the
section labels (e.g., verse, chorus, and bridge).
We also propose a batch sampling scheme to ensure the labels in a training pair are interpreted meaningfully. Given several relevant open-source packages that can help achieve this work, we introduce a modular pipeline
including various metric learning methods and MSA algorithms, and make clear what parts of the system can be easily changed. By using the embeddings as features for an existing MSA algorithm, our supervised approach can support both segmentation and labeling tasks. 
In experiments, we leverage Harmonix Set, SALAMI, and RWC~\cite{goto2002rwc} to investigate the performance in intra- and cross-dataset scenarios.

\section{Related work}

Many MSA approaches (see~\cite{nieto2020audio} for an overview) are supervised in the sense of being tuned to a dataset---e.g., by setting a filter size according to the average segment duration in a corpus. 
An advanced version of this is~\cite{mcfee2014learning}, in which a recurrence matrix is transformed to match the statistics of a training dataset.
However, supervised training has only been used in a few instances for MSA.

The first such method used supervision to learn a notion of `boundaryness' directly from audio~\cite{ullrich2014_ismir}; the method was refined to use a self-similarity lag matrix computed from audio~\cite{grill2015music}.
Similarly,~\cite{wang2021supervised} used supervision to learn what characterizes boundaries as well as ``chorusness'' in audio, and used it in a system to predict the locations of choruses in songs, which is a subproblem of MSA.
Although these 3 systems all have an `end-to-end' flavor, in fact they required the invention of new custom pipelines to obtain estimates of structure, e.g., a peak-picking method to select the likeliest set of boundaries from a boundary probability curve.
The post-processing is also complex in~\cite{maezawa2019music}, in which a boundary fitness estimator similar to~\cite{grill2015music} is combined in a hybrid model with a trained segment length fitness estimator and a hand-crafted timbral homogeneity estimator.
In our work, we aim to arrive at a feature representation that can be used with existing pipelines.

Taking the converse approach,~\cite{shibata2020music} used supervision to model how traditional features (MFCCs, CQT, etc.) relate to music structure, using an LSTM combined with a Hidden semi-Markov Model. Since our approaches are complementary, a combined approach---inputting deep structure features to the LSTM-HSMM---may prove successful, and should be explored in future work.





As noted in the previous section, metric learning was previously applied to improve MSA by~\cite{mccallum2019unsupervised}, but that work took an unsupervised approach: audio fragments in a piece were presumed to belong to the same class if they were near each other in time, and to different classes otherwise. This is a useful heuristic, but by design we expect it to use many false positive and false negative pairs in training.
Also, that work did not report any evaluation on whether the learned embeddings could help with the segment labeling task, nor on the impact of many choices made in the system that could affect the results: the model architecture, loss function, and segmentation method.
In this work, we conduct evaluations on the segmentation and labeling tasks, and investigate the impact of these design choices.
The supervision strategy in this work differs from prior art, and to our knowledge, this work represents the first attempt to develop supervised feature learning with a goal of improving existing MSA algorithms. 


%
%
%
%
%
%
%

\section{System Overview}\label{sec:sys}

The training pipeline of our proposed system is illustrated in Fig. \ref{fig:sys}, and is divided into three stages: (1) feature extraction, (2) mining and training, and (3) validation.

The feature extraction stage consists of two modules. Following most state-of-the-art MSA algorithms~\cite{nieto2020audio}, we synchronize the audio features with beat or downbeats. We use madmom~\cite{Madmom} to estimate the beats and downbeats, and use these to create audio inputs to train a DNN; details of this are explained in Section 4.1. The network outputs the embedding vectors of a song for a subsequent algorithm to complete the task.

The mining and training stage covers four modules: batching, which we define ourselves, followed by miner, loss and distance modules, for which we use PML (pytorch-metric-learning\footnote{https://github.com/KevinMusgrave/pytorch-metric-learning}), an open-source package with implementation options for each.



\textit{Batching}: The training data are split into batches with a fixed size. 
To allow sensible comparisons among the training examples within a batch, we propose a scheme that ensures a batch only contains examples from the same song.

\textit{Miner}: Given the embeddings and labels of examples in a batch, the miner provides an algorithm to pick \textit{informative} training tuples (e.g., a pair having different labels but a large similarity) to compute the loss. Conventional metric learning methods
just use all tuples in a batch (or, sample them uniformly) to train the model. As the batch size grows, using an informative subset can speed up convergence and provide a better model~\cite{wang2019multi}. 

\textit{Loss}: 
PML provides many well-known loss functions developed for deep metric learning, such as contrastive loss~\cite{hadsell2006dimensionality} and triplet loss~\cite{hoffer2015deep}.
We instead use MultiSimilarity loss~\cite{wang2019multi} (see Section \ref{sec:mine_loss}), a more general framework that unifies aspects of multiple weighting schemes
that has not yet been used in an MIR application.

\textit{Distance}: The distance metric defines the geometrical relationship among the output embeddings. Common metrics include Euclidean distance and cosine similarity.

For the validation stage, an MSA algorithm is adopted to generate the boundary and label outputs and validate the model learning status in terms of music structure analysis. The open-source package MSAF has implemented a representative sample of traditional  algorithms~\cite{nieto2016systematic}. 
An algorithm for a different task could be inserted here to tie the training to a different objective.

\section{Technical Details}\label{sec:details}

\subsection{Deep Neural Network Module}\label{subsec:dnn}

The input to the \textit{DNN module} is defined to be a window (e.g. 8 second) of waveform audio, and the output to be a multi-dimensional embedding vector.
We use a two-stage architecture in which the audio is transformed to a time-frequency representation before entering the DNN, but a fully end-to-end DNN would be possible.

In this work, we study two existing two-stage model architectures: Harmonic-CNN~\cite{won2020data} and ResNet-50~\cite{he2016identity}. These open-source architectures have shown good performance in general-purpose music audio classification (e.g., auto-tagging~\cite{won2020data}), so we believe they can be trained to characterize useful information related to music sections. We replace their final two layers (which conventionally consist of a dense layer plus a sigmoid layer) with an \emph{embedding module}, which in turn contains a linear layer, a leaky ReLU, a batch normalization, a linear layer, and a L2-normalization at the output. The input and output dimensions of this module are 256 and 100, respectively.


Any model with a similar purpose could be used for the DNN module in the proposed general framework. We have chosen the above architectures for their convenience, but they could be unsuitably complex given the small size of the available MSA training data. Developing a dedicated, optimal architecture is a task for future work.

\subsection{Audio Input, Alignment, and Label}

\begin{figure}
\centering
\includegraphics[width=\columnwidth]{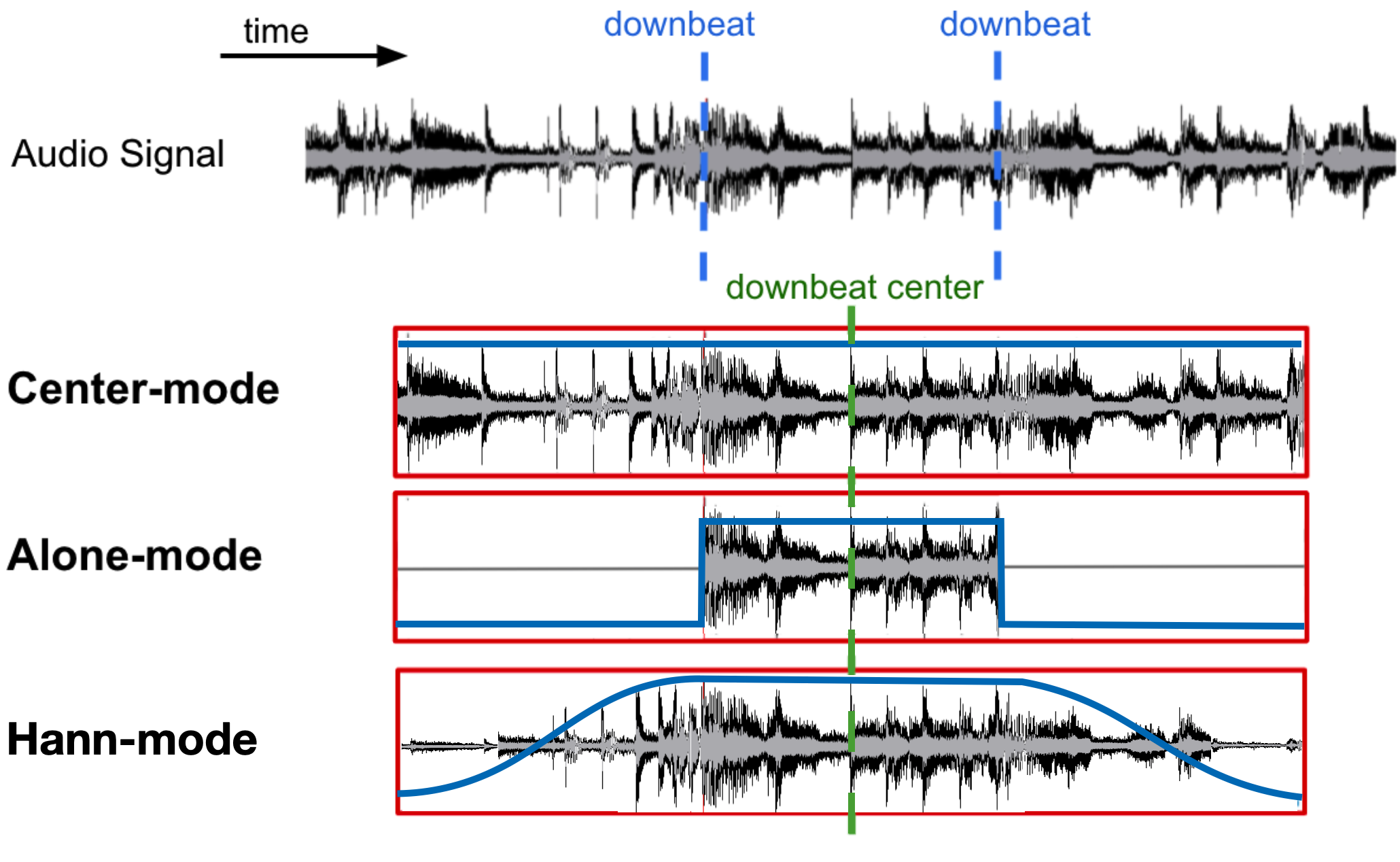}
\caption{Each red box presents a window mode.}
\label{fig:windowing}
\end{figure}


In order to synchronize the output embeddings with downbeats, we align the center of an input window to the center of each bar interval. The same procedure applies if aligning to beats.
Typically, the input window is much longer than the duration of a bar, so there is additional context audio before and after the downbeat interval.
We try three windowing methods that weight this context differently, also as illustrated in Fig. \ref{fig:windowing}:
\emph{center-mode}, where the window is unaltered; \emph{alone-mode}, where the context audio is zeroed out; and \emph{Hann-mode}, where a Hann-shaped ramp from 0 to 1 is applied to the context audio. In our pilot studies, Hann-mode performed best, indicating that some context is useful, but the model should still focus on the signal around the center.

Annotations of structure contain, for each section, two timestamps defining the interval, and a label. These labels may be explicit functions (e.g., intro, verse, chorus) or abstract symbols
(e.g., A, A$'$, B, and C) indicating repetition. A training example is assigned with a label according to the label of the exact center in the input audio. We denote a training example aligned with the $i^{th}$ beat/downbeat of the $j^{th}$ song by $s_i^j = (\mathbf{x}^j_i, y^j_i)$, where $\mathbf{x}$ and $y$ are the audio and label, respectively.

\subsection{Batch Sampling Scheme}

Let a dataset be denoted by $\{[s_i^j]_{i=1}^{m_j}\}_{j=1}^M$, where the $j^{th}$ song has $m_j$ examples. The proposed batch sampling scheme ensures that no cross-song examples are sampled in a batch. Therefore, when comparing any examples within a batch, the labels are meaningful for supervision. For example, we do not want a chorus fragment of song A to be compared with a chorus fragment of song B, since we have no \textit{a priori} way to know whether these should be embedded near or far in the space.

Algorithm \ref{algo:sampling} gives the procedure for one epoch,
i.e., one full pass of the dataset.
We shuffle the original input sequence (line 2) to ensure that each batch is diverse, containing fragments from throughout the song.
Lines 4--6 ensure, when more than one batch is needed for a song, the last batch is full by duplicating examples within the song.
Once a batch is sampled (line 8), we can run a miner to select informative pairs from the batch to calculate the loss to update the model.

\begin{algorithm} [t]
\caption{One epoch of learning procedure.}
\label{algo:sampling}
\SetAlgoNoLine
\LinesNumbered
\DontPrintSemicolon
\KwIn{$\{[s_i^j]_{i=1}^{m_j}\}_{j=1}^M$ , model $\mathbf{\Theta}$, and batch size $\beta$ }
\KwOut{Learned model $\mathbf{\hat\Theta}$}
    \For{$j = 1$ \KwTo $M$}
    {
        $[s_{i'}^j]_{i'=1}^{m_j} \leftarrow \text{shuffle sequence~} [s^j_i]_{i=1}^{m_j}$ \;
    	$n \leftarrow \lceil {m_j / \beta} \rceil$ \tcp*{\small {number of batches}} 
    	
    	\If {$n > 1$}
    	{
    	    $r \leftarrow n \beta - m_j$ \tcp*{\small {space in batch}}
    	    $ [s_{i'}^j]_{i'=1}^{n \beta}  \leftarrow \text{concat~} [s_{i'}^j]_{i'=1}^{m_j}\text{~and~} [s_{i'}^j]_{i'=1}^r$\;
    	}
        \For{$k = 1$ \KwTo $n$}{
    		$\mathcal{B} \leftarrow \{s_{i'}^j\} \text{,~~} i'= \beta(k-1):\text{min}(\beta k, m_j)$ \;
    		$\mathbf{\hat\Theta} \leftarrow \text{update~} \mathbf{\Theta}\text{~with loss computed on~} \mathcal{B}$ \; 
    	}
    }
\end{algorithm}

\begin{figure*}
\centering
\includegraphics[width=\textwidth]{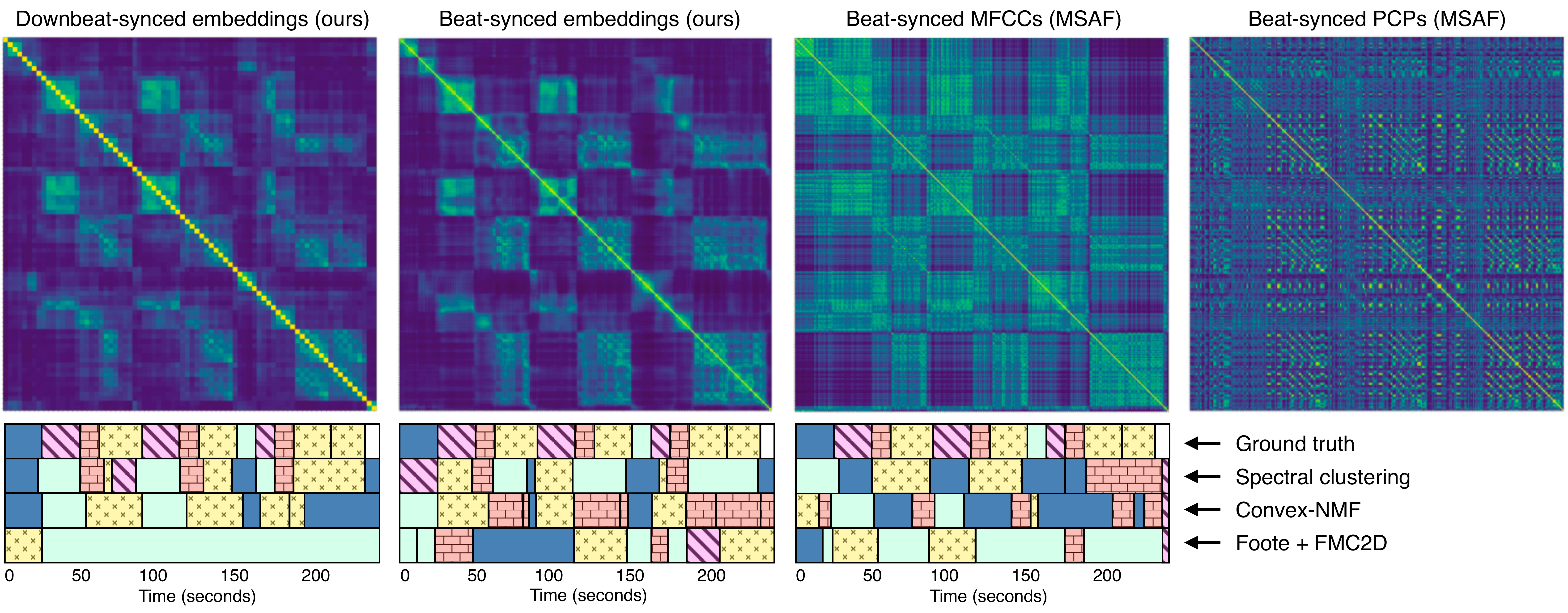}
\caption{Four SSMs using different features for a test song \emph{Avril Lavigne - Complicated}: two versions of the proposed embeddings (left), and two standard features (right). Below the SSMs are the segments and labels for the ground truth analysis, plus the estimated analyses from three algorithms.
The block colors indicate the label clusters within a song. 
}
\label{fig:ssm_comp}
\end{figure*}

\subsection{Miner and Loss} \label{sec:mine_loss}


The MultiSimilarity framework~\cite{wang2019multi} uses three types of similarities to estimate the importance of a potential pair: self-similarity (Sim-S), positive relative similarity (Sim-P), and negative relative similarity (Sim-N).
The authors show that many existing deep metric learning methods only consider one of these types when designing a loss function. By considering all three types of similarities, MultiSimilarity offers stronger capability in weighting important pairs, setting new state-of-the-art performance in image retrieval. From our experiments, it also demonstrates better accuracy over other methods.

For an anchor $s_i^j$, an example $s_k^j$ will lead to a positive pair if they have the same label (i.e., $y_i^j=y_k^j$), and a negative pair otherwise (i.e., $y_i^j\neq y_k^j$).
At the minor phase, the algorithm calculates the Sim-P's for each positive/negative pair against an anchor, and selects the challenging pairs when certain conditions are satisfied.
At the loss phase, it uses the Sim-S's and Sim-N's to calculate the weights for the positive and negative pairs, respectively, where the weights are actually the gradients for updating the model.
To summarize, MultiSimilarity aims to minimize intra-class dissimilarity at the mining stage, and to simultaneously maximize intra-class similarity and minimize inter-class similarity at the loss stage.
In musical terms, the desired result is that fragments with the same section type will be embedded in tight clusters, and that clusters for different sections will be far from one another.

\subsection{MSA Algorithms}

The typical input to an MSA algorithm~\cite{nieto2016systematic} is a sequence of feature vectors. Then, the algorithm outputs the predicted timestamps and an abstract label for each segment.

Fig. \ref{fig:ssm_comp} presents four self-similarity matrices (SSMs) of the same test song using different features. We compute the pairwise Euclidean distance matrix and then apply a Gaussian kernel (see~\cite{mcfee2014analyzing} for details) 
to derive the pairwise similarity. The left two matrices are based on a Harmonic-CNN trained with the MultiSimilarity miner and loss; the right two matrices are based on two traditional features, MFCCs and PCPs.
We see that, compared to traditional features, the learned features can enhance the blocks considerably in the images, reducing the complexity faced by the MSA algorithm.

\begin{figure}
\centering
\includegraphics[width=\columnwidth]{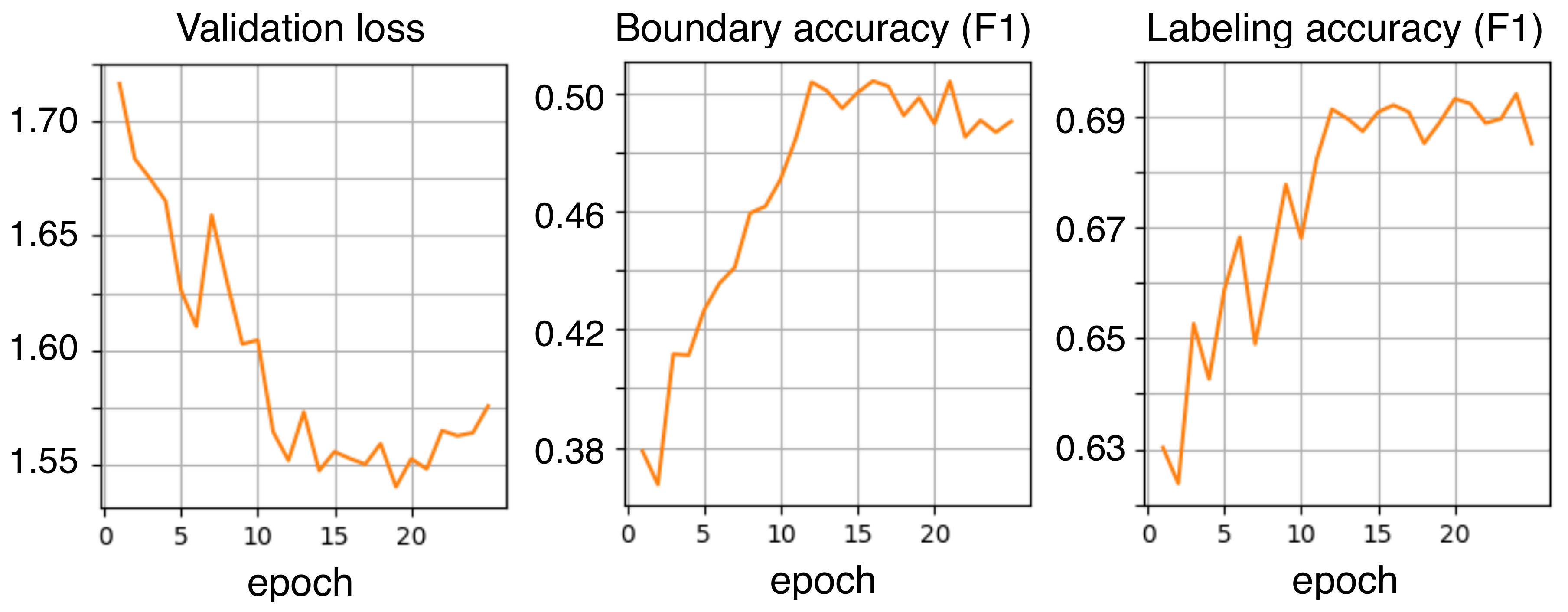}
\caption{Validation loss vs. MSA (scluster) performance.}
\label{fig:valid_curves}
\end{figure}

We picked three MSA algorithms to study here: spectral clustering (\emph{scluster})~\cite{mcfee2014analyzing}, convex-NMF (\emph{cnmf})~\cite{nieto2013convex}, and \emph{foote+fmc2d} (using Foote's algorithm~\cite{foote2000automatic} for segmentation and fmc2d~\cite{nieto2014music} for labeling). Note that each is based on analyzing some version of an SSM.
As these algorithms were developed using traditional features, we need to adjust their default parameters in MSAF to be more suitable for a SSM with prominent but blurry blocks, rather than a sharp but noisy SSM typically treated with a low-pass filter to enhance the block structure.
Also, some MSA algorithms can be sensitive to temporal resolution, which prefers beat- to downbeat-synchronized features. 


How the model training criterion improves an MSA algorithm can be explained in a theoretical way. For example, in scluster, small (e.g., one-bar) fragments of music that have the same labels are considered to be mutually connected in a sub-graph. When the metric learning loss is minimized, scluster is more likely to produce a clear sub-graph for each segment in a song, making the graph decomposition more accurate. As Fig. \ref{fig:valid_curves} illustrates, the evolution of the validation loss is consistent with the performance of scluster when it is fine-tuned to fully exploit the embeddings. This technique provides a guideline to adjust the parameters of an MSA algorithm for most cases.


\section{Experiments}

\subsection{Datasets}

We use three datasets to study the performance: Harmonix Set~\cite{nieto2019harmonix}, SALAMI~\cite{smith2011design}, and RWC (Popular)~\cite{goto2002rwc}.

The Harmonix Set covers a wide range of western popular music, including pop, electronic, hip-hop, rock, country, and metal. Each track is annotated with segment function labels and timestamps.
The original audio to the annotations is not available, but a reference YouTube link is provided. 
We searched for the audio of the right version 
and manually adjusted the annotations to ensure the labels and timestamps were sensible and aligned to the audio.

In SALAMI, some songs are annotated twice; we treat each annotation of a song separately, yielding 2243 annotated songs in total. We also use a subset with 445 annotated songs (from 274 unique songs) in the ``popular'' genre, called  \emph{SALAMI-pop}, for cross-dataset evaluation.

The Popular subset of RWC is composed of 100 tracks. There are two versions of the ground truth: one originally included with the dataset (AIST), and the other provided by INRIA~\cite{bimbot2010decomposition}. The INRIA annotations contain boundaries but not segment labels.

Table \ref{tab:dataset_stats} lists some properties of the datasets. ``Num'' is the number of annotated songs. The number of unique labels per song (``Uni'') ranges between 5.7 and 7.8, indicating that the segment labels are not too repetitive nor too diverse and thus can offer adequate supervision for metric learning. Additional statistics like the number of segments per song (``Seg'') and the mean duration per segment in second (``Dur'') are all within a proper range. Three of the datasets are employed in MIREX, so we can compare our systems with historical ones.

\subsection{Evaluation Metrics}

We focus on flat annotations (i.e. non-hierarchical) in our experiments. The evaluation metrics for MSA have been well-defined, and details can be found in~\cite{nieto2020audio}. We use the following:
(1) \emph{HR.5F}: F-measure of hit rate at 0.5 seconds;
(2) \emph{HR3F}: F-measure of hit rate at 3 seconds;
(3) \emph{PWF}: F-measure of pair-wise frame clustering;
(4) \emph{Sf}: F-measure of normalized entropy score.
Hit rate measures how accurate the predicted boundaries are within a tolerance range (e.g., $\pm$ 0.5 seconds). Pair-wise frame clustering is related to the accuracy of segment labeling. Normalized entropy score gives an estimate about how much a labeling is over- or under-segmented. 

\subsection{Implementation Details}

PyTorch 1.6 is used. We adopt audio windows of length 8 seconds, which we found better than 5 or 10 seconds. 
The audio is resampled at 16KHz and converted to log-scaled mel-spectrograms using 512-point FFTs with 50\% overlap and 128 mel-components. 
We follow~\cite{harmonic-cnn} and~\cite{resnet-50} to implement Harmonic-CNN and ResNet-50, respectively.
For the miner and loss in pytorch-metric-learning, the default parameters suggested by the package are adopted.
We employ the Adam optimizer to train the model, and monitor the \emph{MSA summary score}, defined as
$\frac{5}{14}(\textit{HR.5F}) + \frac{2}{14}(\textit{HR3F}) + \frac{4}{14} (\textit{PWF}) + \frac{3}{14} (\textit{Sf})$, to determine the best model.
The weights were chosen intuitively, but could be optimized in future work.
We use a scheduled learning rate starting at 0.001, and then reduced by 20\% if the score is not increased in two epochs.
We train the models on a Tesla-V100-SXM2-32GB GPU with batch sizes of 128 and 240 for Harmonic-CNN and ResNet-50, respectively.

Regarding fine-tuning the parameters for MSAF, we run a simple grid search using a limited set of integer values on the validation set. As mentioned, the parameters are mostly different from the defaults. For instance, in scluster, we set (``evec\_smooth'', ``rec\_smooth'', ``rec\_width'') as (5, 3, 2), which were (9, 9, 9) by default.
Also, scluster was designed to use separate timbral and harmonic features, but we use the same proposed features for both.

\begin{table}
\small \centering
 \begin{tabular}{l|ccccc}
Dataset & Num & Uni & Seg & Dur & MIREX \\
\hline
Harmonix Set & 912 & 5.7 & 10.6 & 21.7 & \xmark \\
SALAMI & 2243 & 5.9 & 12.5 & 24.2 & \cmark \\
SALAMI-pop & 445 & 6.4 & 13.2 & 18.0 & \xmark \\
RWC-AIST & 100 & 7.8 & 15.3 & 15.2 & \cmark \\
RWC-INRIA & 100 & - & 15.3 & 15.0 & \cmark \\
 \end{tabular}
\caption{Dataset and segment label statistics.}
\label{tab:dataset_stats}
\end{table}

\subsection{Result and Discussion}


\begin{table}
 \small \centering
 \begin{tabular}{l|l|cccc}
\footnotesize Model & \footnotesize System & \footnotesize HR.5F & \footnotesize HR3F & \footnotesize PWF & \footnotesize Sf \\
\hline
\hline
\multirow{3}{*}{Base}
& cnmf/B & .183 & .453 & .498 & .566 \\
& ft+fmc2d/B & .242 & .584 & .536 & .592 \\
& scluster/B & .263 & .547 & .586 & .641 \\
\hline
\multirow{3}{*}{Harm} 
& cnmf/B/eu/mul     & .352 & .679 & .647 & .681 \\
& ft+fmc2d/B/eu/mul & .395 & .713 & .580 & .630 \\
& scluster/B/eu/mul & .466 & .728 & \textbf{.689} & .737 \\
\hline
\multirow{3}{*}{ResN} 
& cnmf/B/eu/mul     & .339 & .637 & .618 & .661 \\
& ft+fmc2d/B/eu/mul & .373 & .685 & .572 & .634 \\
& scluster/D/eu/mul & .433 & .720 & .673 & .728 \\
\hline
\multirow{6}{*}{Harm} 
& scluster/D/eu/mul & \textbf{.497} & \textbf{.738} & .684 & \textbf{.743} \\
& scluster/D/co/mul & .474 & .706 & .668 & .727 \\
& scluster/D/eu/tri & .454 & .713 & .669 & .722 \\
& scluster/D/co/tri & .448 & .693 & .659 & .713 \\
& scluster/D/eu/con & .435 & .682 & .635 & .698 \\
 \end{tabular}
\caption{Cross-validation result on the Harmonix Set. Top 9 rows: Comparison of different models \{`Base': baseline, `Harm': Harmonic-CNN, `ResN': ResNet-50\} and MSA methods at beat-level (`B'). Bottom 6 rows: comparison of different distances \{`eu': Euclidean, `co': cosine\} and losses \{`mul': MultiSimilarity, `tri': TripletMargin, `con': Contrastive\} options, at downbeat-level (`D'). `ft' stands for Foote~\cite{foote2000automatic}.} 
\label{tab:harmonix}
\end{table}

We present three sets of evaluations: (1) a comparison of many versions of our pipeline to establish the impact of the choice of modules; (2) a cross-dataset evaluation; and (3) a comparison of our system with past MIREX submissions.

First, we study the effect of several options for the proposed pipeline: (1) beat or downbeat alignment for input audio; (2) distance metric for the learned features; (3) miner and loss for metric learning. For (3), we use the proposed MultiSimilarity approach and TripletMargin miner and loss~\cite{hoffer2015deep}; we also test Contrastive loss~\cite{hadsell2006dimensionality} with a BaseMiner, which samples pairs uniformly. Each version of the feature embedding is trained and tested on the Harmonix Set using 4-fold cross-validation.

We compare the success of three MSA algorithms when using the proposed features and when using conventional features. In all cases, we synchronize the features to the beats/downbeats estimated by madmom; for the proposed features, we use the ground-truth beats/downbeats for training and the estimated ones for testing.
For a fair comparison, we fine-tune the algorithm parameters for each algorithm-feature combination (including the conventional features) by running a grid search and optimizing the MSA summary score on the training set.

Table \ref{tab:harmonix} presents the results. They show that every MSA algorithm is improved by using the learned features instead of the baseline ones, by a wide margin: HR.5F nearly doubles in most cases when switching to learned features.
The performance differences for each algorithm (e.g., `Base scluster/B' versus `Harm scluster/B/eu/mul') are significant with p-values < $10^{-5}$ for every metric.
The top MSA algorithm overall is scluster, which performs the best on boundary hit rate when synchronized with downbeats, but performs slightly better on PWF when synchronized to beats.
Comparing the two architectures, Harmonic-CNN performs better than ResNet-50 in general, perhaps because the deeper ResNet model requires more data. 

Regarding the other training settings, we find that using Euclidean distance was consistently better than using cosine distance, and that the MultiSimilarity loss gave consistently better results than the other loss functions.
While running the experiments, we notice that 
with Euclidean distance, the validation loss evolved in a more stable way.

\begin{table}
 \small \centering
 \begin{tabular}{l|l|cccc}
\footnotesize Model & \footnotesize System & \footnotesize HR.5F & \footnotesize HR3F & \footnotesize PWF & \footnotesize Sf \\
\hline
\multirow{4}{*}{Base}
& cnmf/B        & .259 & .506 & .485 & .521 \\
& ft+fmc2d/B & .319 & .593 & .521 & .551 \\
& scluster/B    & .305 & .553 & .545 & .572 \\
\hline
\multirow{4}{*}{Harm} 
& cnmf/B/eu/mul     & .301 & .573 & .588 & .601 \\
& ft+fmc2d/B/eu/mul & .358 & .599 & .538 & .581 \\
& scluster/B/eu/mul & .378 & .613 & \textbf{.621} & .644 \\
& scluster/D/eu/mul & \textbf{.447} & \textbf{.623} & .615 & \textbf{.653} \\
 \end{tabular}
\caption{Cross-dataset result on SALAMI-pop (trained on Harmonix Set); and ``ft'' stands for Foote.}
\label{tab:salamipop}
\end{table}

\begin{table}
 \small  
 \begin{tabular}{p{2.2cm}|p{.46cm}p{.46cm}p{.46cm}p{.46cm}|p{.46cm}p{.5cm}} 
 & \multicolumn{4}{c|}{AIST}  &  \multicolumn{2}{c}{INRIA}  \\
 System &  \footnotesize{HR.5F} & \footnotesize{HR3F} & \footnotesize{PWF} & \footnotesize{Sf}  &  \footnotesize{HR.5F} & \footnotesize{HR3F} \\
\hline
OLDA+fmc2d  & .255 & .554 & .526 & .613 & .381 & .604 \\
SMGA2 (2012) & .246 & .700 & .688 & .733 & .252 & .759 \\
GS1 (2015)  & \textbf{.506} & \textbf{.715} & .542 & .692 & \textbf{.697} & \textbf{.793} \\
scluster/D/eu/mul & .438 & .653 & \textbf{.704} & \textbf{.739} & .563 & .677 \\
 \end{tabular}
\caption{MIREX-RWC (cross-dataset) result.}
\label{tab:mirex-rwc}
\end{table}

Second, we study cross-dataset performance by using the best trained model on Harmonix Set to make predictions for the songs in SALAMI-pop.
This tests the model ability to avoid overfitting to one style of annotations.
In Table \ref{tab:salamipop}, we see that the scluster algorithm again performs the best, and again improves significantly when using the learned features (p-value < $10^{-10}$). However, the improvement margins are smaller for cnmf and foote+fmc2d (e.g., for cnmf, HR.5F increases by $0.042$; before, it increased by $0.169$).
Perhaps the MSA parameters for these two algorithms are over-tuned to the training data; or, it may be that the learned features overfit the style of pop in Harmonix Set, but that scluster is more robust to this.


Finally, we collect previous MIREX results to compare our system to others.
For the RWC (popular) task, we use the same model (trained on Harmonix Set) from the previous experiment on SALAMI-pop. The results are shown in Table \ref{tab:mirex-rwc} alongside those of three of the strongest MIREX submissions. We omit some, like SMGA1, that are related to or based on the same approach as others listed but perform worse.
Our system can outperform the state-of-the-art (SMGA2) in terms of PWF and Sf. Regarding its segmentation performance, it is still competitive, outperforming OLDA (the top-performing segmenter offered by MSAF) by a large margin (HR.5F/3F).

For the SALAMI task, the identity of the songs used in MIREX is private, but 487 songs (with 921 annotations) have been identified~\cite{salami_mirex}. We use this portion as the test set, and the remainder of SALAMI (1322 annotations of 872 songs) as the sole training and validation set. The results are shown in Table \ref{tab:mirex-salami}, along with other MIREX competitors, including OLDA+FMC2D, SMGA1, and GS1 (which uses a CNN trained to directly model the boundaries~\cite{grill2015music}).
As SALAMI is more diverse than Harmonix Set, the model sees fewer examples per style compared to when it was trained on Harmonix Set. Thereby, we can expect the learned features to be less successful.
However, we once again see that each model in MSAF is improved on all metrics when using the learned features, particularly in terms of PWF. In fact, our model boosts cnmf---already third-best among the baseline algorithms shown here---to outperform the state-of-the-art (SMGA1).

\begin{table}
 \small \centering
 \begin{tabular}{l|cccc}
 System &  HR.5F & HR3F & PWF & Sf \\
\hline
cnmf (2016)       & .228 & .427 & .527 & .543 \\
foote+fmc2d (2016) & .244 & .503 & .463 & .549 \\
scluster (2016) & .255 & .420 & .472 & .608 \\
OLDA+fmc2d~\cite{mcfee2014learning,nieto2016mirex}  & .299 & .486 & .471 & .559 \\
SMGA1 (2012)~\cite{serra2012importance} & .192 & .492 & .581 & .648 \\
Segmentino~\cite{matthias2009a,cannam2016mirex} & .209 & .475 & .551 & .612 \\
GS1 (2015)~\cite{grill2015structural,grill2015music}  & \textbf{.541} & \textbf{.623} & .505 & \textbf{.650} \\
\hline
cnmf/D/eu/mul     & .318 & .506 & \textbf{.587} & .578 \\
foote+fmc2d/B/eu/mul & .289 & .519 & .558 & .563 \\
scluster/D/eu/mul & .356 & .553 & .568 & .613 \\
 \end{tabular}
\caption{MIREX-SALAMI result.}
\label{tab:mirex-salami}
\end{table}


The MSAF algorithms are improved with the learned features, but they still lag behind GS1. This is reasonable, since the training of that model directly connects to the loss of boundary prediction, and ours does not.
Nonetheless, ``scluster/D/eu/mul'' can outperform all the other systems except GS1 by a large margin on both HR.5F and HR3F.




\section{Conclusion and future work}

We have presented a modular training pipeline to learn the deep structure features for music audio. The pipeline consists of audio pre-processing, DNN model, metric learning module, and MSA algorithm. We have explained the functionality for each component and demonstrated the effectiveness of different module combinations. In experiments, we have found that using the learned features can improve an MSA algorithm significantly.

However, the model is not yet fully end-to-end: the MSA outputs (boundaries and labels) are not directly back-propagated to the DNN model. We plan to explore ways to change this in future work---e.g., by exploring self-attention models like the Transformer~\cite{vaswani2017attention,lu2021spectnt} to build a deep model that directly outputs the segment clusters.
This would eliminate the need to fine-tune MSA parameters.

\bibliography{citations}

\begin{thebibliography}{10}
\providecommand{\url}[1]{#1}
\csname url@samestyle\endcsname
\providecommand{\newblock}{\relax}
\providecommand{\bibinfo}[2]{#2}
\providecommand{\BIBentrySTDinterwordspacing}{\spaceskip=0pt\relax}
\providecommand{\BIBentryALTinterwordstretchfactor}{4}
\providecommand{\BIBentryALTinterwordspacing}{\spaceskip=\fontdimen2\font plus
\BIBentryALTinterwordstretchfactor\fontdimen3\font minus
  \fontdimen4\font\relax}
\providecommand{\BIBforeignlanguage}[2]{{%
\expandafter\ifx\csname l@#1\endcsname\relax
\typeout{** WARNING: IEEEtran.bst: No hyphenation pattern has been}%
\typeout{** loaded for the language `#1'. Using the pattern for}%
\typeout{** the default language instead.}%
\else
\language=\csname l@#1\endcsname
\fi
#2}}
\providecommand{\BIBdecl}{\relax}
\BIBdecl

\bibitem{mcfee2014analyzing}
B.~McFee and D.~Ellis, ``Analyzing song structure with spectral clustering,''
  in \emph{ISMIR}, 2014, pp. 405--410.

\bibitem{shibata2020music}
G.~Shibata, R.~Nishikimi, and K.~Yoshii, ``Music structure analysis based on an
  {LSTM-HSMM} hybrid model,'' in \emph{ISMIR}, 2020, pp. 15--22.

\bibitem{wang2016structural}
C.~Wang and G.~J. Mysore, ``Structural segmentation with the variable {M}arkov
  oracle and boundary adjustment,'' in \emph{Proc. ICASSP}, 2016, pp. 291--295.

\bibitem{bruderer2009perception}
M.~J. Bruderer, M.~F. Mckinney, and A.~Kohlrausch, ``The perception of
  structural boundaries in melody lines of western popular music,''
  \emph{Musicae Scientiae}, vol.~13, no.~2, pp. 273--313, 2009.

\bibitem{deberardinis2020unveiling}
J.~de~Berardinis, M.~Vamvakaris, A.~Cangelosi, and E.~Coutinho, ``Unveiling the
  hierarchical structure of music by multi-resolution community detection,''
  \emph{Trans. ISMIR}, vol.~3, no.~1, pp. 82--97, 2020.

\bibitem{jehan2005a}
T.~Jehan, ``Hierarchical multi-class self similarities,'' in \emph{Proceedings
  of IEEE Workshop on Applications of Signal Processing to Audio and
  Acoustics}, 2005, pp. 311--314.

\bibitem{ullrich2014_ismir}
K.~Ullrich, J.~Schl{\"u}ter, and T.~Grill, ``{Boundary detection in music
  structure analysis using convolutional neural networks},'' in \emph{ISMIR},
  2014, pp. 417--422.

\bibitem{wang2021supervised}
J.-C. Wang, J.~B.~L. Smith, J.~Chen, X.~Song, and Y.~Wang, ``Supervised chorus
  detection for popular music using convolutional neural network and multi-task
  learning,'' in \emph{Proc. ICASSP}, 2021, pp. 566--570.

\bibitem{mccallum2019unsupervised}
M.~C. McCallum, ``Unsupervised learning of deep features for music
  segmentation,'' in \emph{Proc. ICASSP}, 2019, pp. 346--350.

\bibitem{nieto2019harmonix}
O.~Nieto, M.~McCallum, M.~Davies, A.~Robertson, A.~Stark, and E.~Egozy, ``{The
  Harmonix Set: Beats, downbeats, and functional segment annotations of western
  popular music},'' in \emph{ISMIR}, 2019, pp. 565--572.

\bibitem{smith2011design}
J.~B.~L. Smith, J.~A. Burgoyne, I.~Fujinaga, D.~D.~Roure, and J.~S. Downie,
  ``Design and creation of a large-scale database of structural annotations,''
  in \emph{ISMIR}, 2011, pp. 555--560.

\bibitem{goto2002rwc}
M.~Goto, H.~Hashiguchi, T.~Nishimura, and R.~Oka, ``{RWC Music Database}:
  Popular, classical and jazz music databases.'' in \emph{ISMIR}, 2002, pp.
  287--288.

\bibitem{nieto2020audio}
O.~Nieto, G.~J. Mysore, C.-i. Wang, J.~B.~L. Smith, J.~Schl{\"u}ter, T.~Grill,
  and B.~McFee, ``Audio-based music structure analysis: Current trends, open
  challenges, and applications,'' \emph{Trans. ISMIR}, vol.~3, no.~1, 2020.

\bibitem{mcfee2014learning}
B.~McFee and D.~P. Ellis, ``Learning to segment songs with ordinal linear
  discriminant analysis,'' in \emph{Proc. ICASSP}, 2014, pp. 5197--5201.

\bibitem{grill2015music}
T.~Grill and J.~Schl{\"u}ter, ``Music boundary detection using neural networks
  on combined features and two-level annotations.'' in \emph{ISMIR}, 2015, pp.
  531--537.

\bibitem{maezawa2019music}
A.~Maezawa, ``Music boundary detection based on a hybrid deep model of novelty,
  homogeneity, repetition and duration,'' in \emph{Proc. ICASSP}, 2019, pp.
  206--210.

\bibitem{Madmom}
S.~B{\"o}ck, F.~Korzeniowski, J.~Schl{\"u}ter, F.~Krebs, and G.~Widmer,
  ``{madmom: a new Python Audio and Music Signal Processing Library},'' in
  \emph{Proceedings of the ACM International Conference on Multimedia}, 2016,
  pp. 1174--1178.

\bibitem{wang2019multi}
X.~Wang, X.~Han, W.~Huang, D.~Dong, and M.~R. Scott, ``Multi-similarity loss
  with general pair weighting for deep metric learning,'' in \emph{Proceedings
  of IEEE Conference on Computer Vision and Pattern Recognition}, 2019, pp.
  5022--5030.

\bibitem{hadsell2006dimensionality}
R.~Hadsell, S.~Chopra, and Y.~LeCun, ``Dimensionality reduction by learning an
  invariant mapping,'' in \emph{Proceedings of the IEEE Conference on Computer
  Vision and Pattern Recognition}, 2006, pp. 1735--1742.

\bibitem{hoffer2015deep}
E.~Hoffer and N.~Ailon, ``Deep metric learning using triplet network,'' in
  \emph{International workshop on similarity-based pattern recognition}.\hskip
  1em plus 0.5em minus 0.4em\relax Springer, 2015, pp. 84--92.

\bibitem{nieto2016systematic}
O.~Nieto and J.~P. Bello, ``Systematic exploration of computational music
  structure research,'' in \emph{ISMIR}, 2016, pp. 547--553.

\bibitem{won2020data}
M.~Won, S.~Chun, O.~Nieto, and X.~Serra, ``Data-driven harmonic filters for
  audio representation learning,'' in \emph{Proc. ICASSP}, 2020, pp. 536--540.

\bibitem{he2016identity}
K.~He, X.~Zhang, S.~Ren, and J.~Sun, ``Identity mappings in deep residual
  networks,'' in \emph{European conference on computer vision}, 2016, pp.
  630--645.

\bibitem{nieto2013convex}
O.~Nieto and T.~Jehan, ``Convex non-negative matrix factorization for automatic
  music structure identification,'' in \emph{Proc. ICASSP}, 2013, pp. 236--240.

\bibitem{foote2000automatic}
J.~Foote, ``Automatic audio segmentation using a measure of audio novelty,'' in
  \emph{Proceedings of the IEEE International Conference on Multimedia and
  Expo}, 2000, pp. 452--455.

\bibitem{nieto2014music}
O.~Nieto and J.~P. Bello, ``Music segment similarity using {2D-Fourier}
  magnitude coefficients,'' in \emph{Proc. ICASSP}, 2014, pp. 664--668.

\bibitem{bimbot2010decomposition}
F.~Bimbot, O.~Le~Blouch, G.~Sargent, and E.~Vincent, ``Decomposition into
  autonomous and comparable blocks: a structural description of music pieces,''
  2010.

\bibitem{harmonic-cnn}
M.~Won,
  https://github.com/minzwon/sota-music-tagging-models/blob/master/training/model.py,
  Last accessed on May 5, 2021.

\bibitem{resnet-50}
TorchVision, https://github.com/pytorch/vision/blob/
  master/torchvision/models/resnet.py, Last accessed on May 5, 2021.

\bibitem{salami_mirex}
J.~Schl{\"u}ter, http://www.ofai.at/research/impml/projects/
  audiostreams/ismir2014/salami\_ids.txt, Last accessed on May 5, 2021.

\bibitem{nieto2016mirex}
O.~Nieto, ``{MIREX: MSAF} v0.1.0 submission,'' \emph{Proceedings of the Music
  Information Retrieval Evaluation eXchange (MIREX)}, 2016.

\bibitem{serra2012importance}
J.~Serra \emph{et~al.}, ``The importance of detecting boundaries in music
  structure annotation,'' \emph{Proceedings of the Music Information Retrieval
  Evaluation eXchange (MIREX)}, 2012.

\bibitem{matthias2009a}
M.~Mauch, K.~C. Noland, and S.~Dixon, ``Using musical structure to enhance
  automatic chord transcription,'' in \emph{ISMIR}, 2009, pp. 231--236.

\bibitem{cannam2016mirex}
C.~Cannam, E.~Benetos, M.~Mauch, M.~E. Davies, S.~Dixon, C.~Landone, K.~Noland,
  and D.~Stowell, ``{MIREX} 2016: Vamp plugins from the {Centre for Digital
  Music},'' \emph{Proceedings of the Music Information Retrieval Evaluation
  eXchange (MIREX)}, 2016.

\bibitem{grill2015structural}
T.~Grill and J.~Schl{\"u}ter, ``Structural segmentation with convolutional
  neural networks mirex submission,'' \emph{Proceedings of the Music
  Information Retrieval Evaluation eXchange (MIREX)}, p.~3, 2015.

\bibitem{vaswani2017attention}
A.~Vaswani, N.~Shazeer, N.~Parmar, J.~Uszkoreit, L.~Jones, A.~N. Gomez,
  L.~Kaiser, and I.~Polosukhin, ``Attention is all you need,'' \emph{arXiv
  preprint arXiv:1706.03762}, 2017.

\bibitem{lu2021spectnt}
W.-T. Lu, J.-C. Wang, M.~Won, K.~Choi, and X.~Song, ``Spec{TNT}: a
  time-frequency transformer for music audio,'' in \emph{ISMIR}, 2021.

\end{thebibliography}
\end{document}